\documentclass[preprint,12pt]{elsarticle}
\journal{arXiv}

\usepackage[dvipsnames]{xcolor}
\usepackage{amssymb}
\usepackage{amsmath}
\usepackage{enumitem}
\usepackage{bm}
\usepackage{subcaption}
\usepackage{multirow}
\usepackage{array}
\usepackage{booktabs}
\usepackage{url}

\usepackage{newtxtext}
\usepackage[subscriptcorrection]{newtxmath}

\graphicspath{{./image/}}

\usepackage[plain,noend]{algorithm2e}

\newtheorem{theorem}{Theorem}

\DeclareMathOperator*{\argmax}{arg\,max}

\makeatletter
\renewcommand{\algocf@captiontext}[2]{#1\algocf@typo. \AlCapFnt{}#2} 
\def\@algocf@capt@plain{top}
\renewcommand{\algocf@makecaption}[2]{%
  \addtolength{\hsize}{\algomargin}%
  \sbox\@tempboxa{\algocf@captiontext{#1}{#2}}%
  \ifdim\wd\@tempboxa >\hsize
    \hskip .5\algomargin%
    \parbox[t]{\hsize}{\algocf@captiontext{#1}{#2}}
  \else%
    \global\@minipagefalse%
    \hbox to\hsize{\box\@tempboxa}
  \fi%
  \addtolength{\hsize}{-\algomargin}%
}
\makeatother

\begin{document}

\begin{frontmatter}


\title{Bayesian online collective anomaly and change point detection in fine-grained time series}



\author[a]{Xian Chen}
\address[a]{Department of Management Science and Engineering, Shanghai University}

\author[b]{Weichi Wu \corref{cor1}}
\address[b]{Department of Statistics and Data Science, Tsinghua University}
\cortext[cor1]{Corresponding author: wuweichi@mail.tsinghua.edu.cn}


\begin{abstract}
Fine-grained time series data are crucial for accurate and timely online change detection. While both collective anomalies and change points can coexist in such data, their joint online detection has received limited attention. In this research, we develop a Bayesian framework capturing time series with collective anomalies and change points, and introduce a recursive online inference algorithm to detect the most recent collective anomaly and change point jointly. For scaling, we further propose an algorithm enhanced with collective anomaly removal that effectively reduces the time and space complexity to linear. We demonstrate the effectiveness of our approach via extensive experiments on simulated data and two real-world applications.
\end{abstract}

\begin{keyword}
online change detection \sep collective anomalies \sep change points \sep Bayesian inference



\end{keyword}

\end{frontmatter}



\section{Introduction}

Advancements in modern technology have enabled real-time collection of fine-grained time series data, facilitating efficient and effective online change detection \citep{thottan2003anomaly,andreou2004impact} for system monitoring and control in various applications \citep{xie2021sequential}. These \emph{fine-grained} (or high-frequency) time series data consist of observations collected at a high temporal resolution, providing a precise depiction of dynamic processes \citep{nason2017should}. Such data may contain \emph{anomalies} and \emph{change points}, which are defined respectively as temporary abnormal deviations and persistent structural shifts in the data distribution \citep{wendelberger2021monitoring}. These changes have attracted attention in various research domains such as finance and healthcare \citep{chen2000parametric,hilal2022financial}. In transportation, for instance, traffic accidents and infrastructure construction can cause anomalies and change points in travel demand, both of which require timely detection to enhance system resilience \citep{chen2020detecting}.

In fine-grained time series, anomalies usually occur as \emph{collective anomalies} rather than as \emph{point anomalies}. Differentiated by duration, the former are anomalous subsequences spanning several consecutive time points, while the latter occur isolatedly at single time points \citep{chandola2009anomaly}. By adopting a flexible distributional family, both collective anomalies and change points can be formulated by \citep{fisch2022linear, chib1998estimation}:
\begin{equation}
    \theta^t=
    \left\{
    \begin{aligned}
    &\theta_1,\ &\text{if}\ t\leq \tau_1,\\
    &\theta_2,\ &\text{if}\ \tau_1<t\leq \tau_2,\\
    &\vdots&\vdots\\
    &\theta_{K+1},\ &\text{if}\ \tau_K<t\leq T,\\
    \end{aligned}
    \right.
\nonumber
\end{equation}
where $\theta^t\in \{\theta_1,\theta_2,\dots,\theta_{K+1}\}$ is the data distribution parameter at time $t$, $K$ is the number of changes observed by time $T$, time points $1<\tau_1<\tau_2<\dots<\tau_K<T$ indicate either change points or the start and end points of collective anomalies, and a collective anomaly between $\tau_k$ and $\tau_{k+1}$ is distinguished from change points by its short duration $\tau_{k+1}-\tau_k$ and reversible change $\theta_k=\theta_{k+2}\neq \theta_{k+1}$.

\begin{figure}[h]
\centering
\includegraphics[width=0.78\textwidth]{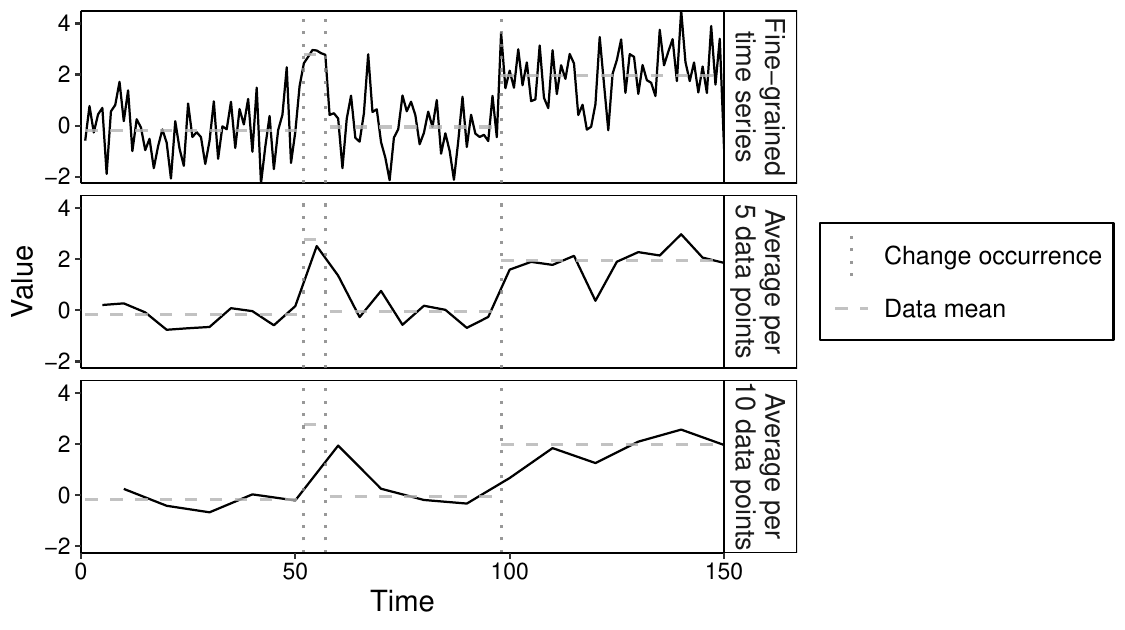}
\caption{An example of time series at different granularities.}
\label{fig:ex_an_cp}
\end{figure}

Fine-grained time series data preserve detailed variation structures and enable more effective detection of collective anomalies and change points, as illustrated in Figure \ref{fig:ex_an_cp}.
The top subplot describes a simulated fine-grained time series containing a collective anomaly from time 52 to 57 and a change point at time 98.
The mean shifts are indicated by grey vertical and horizontal lines. The middle and bottom subplots present the same series averaged over every 5 and 10 consecutive data points, respectively. 
These coarse-grained series smooth out the anomalous fluctuations at the collective anomaly and transform the abrupt change point into a gradual shift, thereby reducing the detectability of these changes.



Although extensive studies have developed online anomaly and change point detection methods, most, if not all, of them focus on point anomalies rather than collective anomalies. Most related to our research is \cite{wendelberger2021monitoring}, which treated point anomalies as generating from a predefined normal distribution, and extend the Bayesian online change point detection algorithm \citep{adams2007bayesian} to infer the most recent point anomaly and change point. 
Researchers in this area have extended anomaly detection methods \citep{takeuchi2006unifying, chen2016general} and explored ensemble techniques \citep{dash2024outlier}. Other remotely related areas include robust online change point detection \citep{fearnhead2019changepoint,knoblauch2018doubly,dehling2020robust,altamirano2023robust}, online collective and point anomaly detection \citep{fisch2022real,tveten2022scalable}, offline point anomaly and change point detection \citep{wu2024trend, tsay1988outliers, chen1993joint}. While methods developed in these areas can identify collective anomalies as consecutive point anomalies, they do not utilize temporal continuity effectively, which can potentially reduce detection performance.

In this paper, we propose to develop a unified framework that jointly detects both collective anomalies and change points in an online manner. This is inherently challenging, as it requires balancing rapid response to changes, accurate classification with sufficient data, and computational efficiency. Specifically, we explore a unified framework, as sequentially detecting these two types of changes without accounting for their potential interactions is problematic: Preprocessing data to eliminate collective anomalies first is challenging in the online scenario \citep{dash2024outlier,song2021sequential}, while detecting change points first often leads to an overestimation of changes \citep{romano2022detecting}. Moreover, although a collective anomaly can be viewed as a data segment bounded by two change points, most online change point detection methods focus on identifying a single change point \citep{xie2021sequential} and thus fail to address our problem. 


Our method builds upon the Bayesian online change point detection (BOCPD) method of \cite{adams2007bayesian} by integrating the classification of recent changes. At each time point, BOCPD updates the posterior estimate of the run length, which is the duration since the most recent change point. The difference between our method and BOCPD mainly lies in two aspects: (1) As BOCPD focuses on only the most recent change, it cannot determine whether two nearby changes detected at different times represent a collective anomaly or refined localization of one change point. To address this issue, we jointly identify two recent changes that correspond to the most recent collective anomaly and change point, respectively; and (2) BOCPD models the prior change probability as a function of the run length. To describe different frequencies of collective anomalies and change points, we extend this function to include the timing and type of the most recent change.

In our framework, we develop a Bayesian model that assigns two binary variables to each time point to indicate whether a collective anomaly or a change point occurs. Specifically, their prior probabilities depend on the timing and type of previous changes. Based on this model, we construct a recursive algorithm to update the posterior estimates for the most recent collective anomaly and change point at each time point. Its time and space complexity is quadratic in the length of the search time range. Hence, we further develop a second algorithm enhanced with collective anomaly removal, effectively reducing the complexity to linear. We discuss hyperparameter selection both theoretically and empirically. Our simulation study shows that our method achieves high accuracy and low delay in detecting collective anomalies and change points. More importantly, our method significantly reduces false alarm rates compared with multiple baseline methods. Two real-world applications on search interest data and electric load data also demonstrate the effectiveness of our method.

 
This paper is organized as follows. Section \ref{sec:model} presents a Bayesian model for time series with collective anomalies and change points. Section \ref{sec:algorithm1} describes a recursive inference algorithm that updates the posterior estimates for the most recent changes. Section \ref{sec:algorithm2} introduces another algorithm with reduced time and space complexity. Section \ref{sec:hyperparameter} discusses hyperparameter selection. Section \ref{sec:simulation} validates our method through simulations, and Section \ref{sec:application} applies it to real-world data. Finally, Section \ref{sec:discussion} concludes with a discussion.

\section{Bayesian modeling}
\label{sec:model}
	
Our detection method is based on a Bayesian modeling framework that captures time series with collective anomalies and change points, both referred to as \emph{changes}. The following notation is used. We apply hyperparameter $\Delta t$ to represent the maximum length of a collective anomaly. We denote the data observed at time $t$ as $y^t$, and assume that it follows a distribution parameterized by $\theta^t$. We use superscript $t':t$ to represent the sequence of variables spanning from time $t'$ to time $t$. Specifically, it denotes an empty set of variables if $t'>t$.

We describe the occurrences of collective anomalies and change points using two sets of binary variables, $\{c^t\}$ and $\{a^t\}$. The former indicates the occurrence of changes, where $c^t=1$ indicates a change at time $t$, and $c^t=0$ otherwise. The latter differentiates between collective anomalies and change points, where $a^t=1$ if a change at time $t$ marks the end of a collective anomaly. That is, the collective anomaly lasts through time $t-1$, and the data distribution reverts to normal at time $t$. Otherwise, $a^t=0$ indicates that the change at time $t$ is either a change point or the start of a collective anomaly. As the classification of changes involves duration threshold $\Delta t$, $c^t$ and $a^t$ are assumed to partially depend on $\bm{c}^{(t-\Delta t):(t-1)}$ and $\bm{a}^{(t-\Delta t):(t-1)}$.

\begin{figure}[h]
\centering
\includegraphics[width=0.85\textwidth]{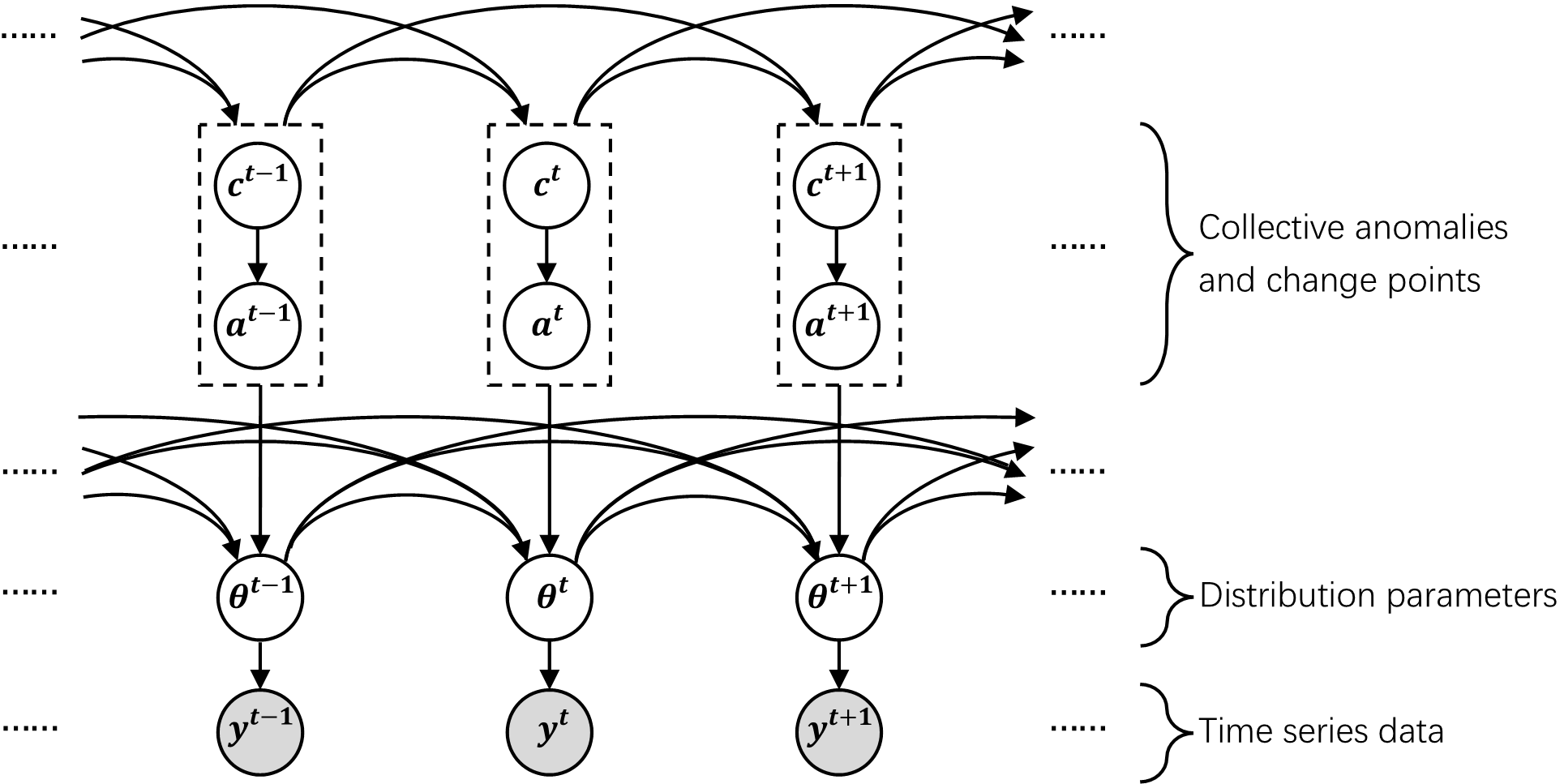}
\caption{The graphical representation of our model structure. The non-shaded and shaded circles represent latent and observed variables, respectively. The arrows denote the dependencies between variables or variable sets.}
\label{fig:proposed_structure}
\end{figure}

We construct a Bayesian generative model capturing the dependencies among variables $\{c^t\}$, $\{a^t\}$, $\{\theta^t\}$, and $\{y^t\}$. Without loss of generality, the first time point of the entire study period is considered a change point. We assume the following data generation process:
\begin{itemize}[itemsep=0pt]
\item For each time point $t=1,\dots,T$:
\begin{itemize}[itemsep=0pt,topsep=0pt]
    \item Sample $c^t\in\{0,1\}$ and $a^t\in\{0,1\}$ from $\mathrm{Pr}\left(c^t,a^t\left|\bm{c}^{\left(t-\Delta t\right):\left(t-1\right)},\bm{a}^{\left(t-\Delta t\right):\left(t-1\right)}\right.\right)$;
    \item Set $\theta^t$ conditional on $c^t$ and $a^t$:
    \begin{itemize}[itemsep=0pt,topsep=0pt]
        \item If $c^t=0$, set $\theta^t=\theta^{t-1}$;
        \item If $c^t=1$ and $a^t=1$, set $\theta^t=\theta^{t'}$ with $t' = \max \{ \tau \mid \tau < t, \, c^{\tau} = 1\}-1$;
        \item Otherwise, sample $\theta^t$ from prior distribution $\mathrm{Pr}\left(\theta^t\left|\pi_0\right.\right)$;
    \end{itemize}
    \item Sample $y^t$ from data distribution $\mathrm{Pr}\left(y^t\left|\theta^t\right.\right)$.
\end{itemize}
\end{itemize}
Here, $\pi_0$ is the hyperparameter in the prior distribution of $\{\theta^t\}$. In accordance with the above process, Figure \ref{fig:proposed_structure} illustrates our model structure with $\Delta t=2$. 

\begin{figure}[h]
\centering
\includegraphics[width=0.6\textwidth]{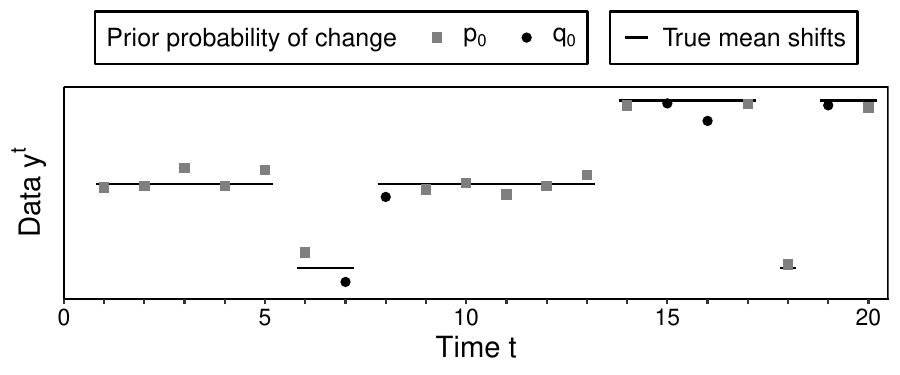}
\caption{Prior change probabilities with $\Delta t=2$ in a simulated time series.}
\label{fig:ex_transit_p}
\end{figure}

We now detail the probabilistic setting when generating the data at time $t$. The choice of data distribution $\mathrm{Pr}(y^t|\theta^t)$ and its prior distribution $\mathrm{Pr}(\theta^t|\pi_0)$ depends on the application context. Additionally, they must be flexible enough to model both normal segments divided by change points and abnormal segments defined as collective anomalies. Let us focus on the generation of $c^t$ and $a^t$. We initialize $c^1=1$ and $a^1=0$ at time $t=1$. At time $t>1$, we set the prior change probability to be a function of the timing and type of the most recent change:
\begin{equation}
    \mathrm{Pr}\left(c^t=1\left|\bm{c}^{\left(t-\Delta t\right):\left(t-1\right)},\bm{a}^{\left(t-\Delta t\right):\left(t-1\right)}\right.\right)=
    \left\{
    \begin{aligned}
    &q_0,\ \text{if}\ \exists\, t'\in \{t-\Delta t,\dots,t-1\}\setminus\{1\}\\
    &\quad \quad s.t.\ c^{t'}=1,\ a^{t'}=0,\  \bm{c}^{\left(t'+1\right):\left(t-1\right)}=\bm{0},\\
    &p_0,\ \text{otherwise},\\
    \end{aligned}
    \right.
    \nonumber
\end{equation}
where $p_0$ is the prior probability that a change of unknown type occurs, and $q_0$ is the prior probability that the previous change is the start of a collective anomaly and the current change marks its end. Using data sampled from a mean-shift normal distribution, Figure \ref{fig:ex_transit_p} presents an example of prior change probabilities with $\Delta t=2$, where grey squares and black circles represent $p_0$ and $q_0$, respectively. Given $c^t$, $a^t$ is then determined by:
\begin{equation}
\begin{aligned}
    a^t=&
    \left\{
    \begin{aligned}
    &1,\ \text{if}\ c^t=1\ \text{and}\ \exists\, t'\in \{t-\Delta t,\dots,t-1\}\setminus\{1\}\ s.t.\ c^{t'}=1,\ a^{t'}=0,\  \bm{c}^{\left(t'+1\right):\left(t-1\right)}=\bm{0},\\
    &0,\ \text{otherwise}.\\
    \end{aligned}
    \right.
\end{aligned}
\nonumber
\end{equation}


\section{Online inference algorithm}
\label{sec:algorithm1}

Based on our model, we develop a Bayesian online change detection (BOCD) algorithm to identify both collective anomalies and change points. In the following, we first describe the algorithmic setup. Then, we detail the calculation of data likelihoods and posterior probabilities used in detection. The supplementary material provides an overview of our algorithm and an analysis of its time and space complexity, which both grow quadratically with the length of the search time range. While our algorithm relies on maximum a posteriori (MAP) estimates, it also provides posterior distributions to quantify the uncertainty in detection results.

\subsection{Algorithmic setup}
\label{subsec:algorithm1_setup}

First, let us denote the search time ranges for collective anomalies and change points at time $t$ as $\Upsilon_a^t$ and $\Upsilon_c^t$, respectively. We impose upper limits $u_a+1$ and $u_c+1$ on their lengths. For clarity, we set all the time indexes as $\{1,\dots,t\}$, those in $\Upsilon_a^t$ as $\{t-n_a^t,\dots,t\}$, and those in $\Upsilon_c^t$ as $\{t-n_c^t,\dots,t\}$. Both $n_a^t$ and $n_c^t$ are set much larger than $\Delta t$, and $n_a^t$ is generally smaller than $n_c^t$. 

Let $r^t$ be the run length variable for the duration from the most recent change to time $t$. We have $r^t=r$ if $c^{t-r}=1$ and $\bm{c}^{(t-r+1):t}=\bm{0}$. Similarly, let variable $d^t$ denote the duration from the most recent change point to time $t$. 
We have $d_t=r^t$ if the most recent change is a change point, and $d^t>r^t$ otherwise. 
If $d^t>r^t$, the change point at time $t-d^t$ should occur at time $1$ or at least $\Delta t+1$ time points before the change at time $t-r^t$. 
To formalize this requirement, let $R(t,d)$ denote the maximum allowed value of $r^t$ satisfying $r^t<d^t$ given $d^t=d$:
\begin{equation}
    R\left(t,d\right)=
    \left\{
    \begin{aligned}
    &d-1,\ \text{if}\ d=t-1,\\
    &d-\Delta t-1,\ \text{otherwise}.\\
    \end{aligned}
    \right.
    \nonumber
\end{equation}
In addition, we consider the case that the most recent collective anomaly ends at time $t$. Its start time $t-1-r^{t-1}$ should be within $\Delta t$ time units of time $t$ and cannot form a collective anomaly with the change point at time $t-d^t$. Here, we denote $A(t,d)$ as the maximum allowed value of $r^{t-1}$ given $r^t=0$, $a^t=1$, and $d^t=d>2$. It takes the value $min\{\Delta t-1, R(t,d)-1\}$, and it is the maximum allowed value of $r^{t-1}$ given $r^t=0$ and $a^t=1$.



It is worth noting that our algorithm does not exactly aligh with our model in three aspects: (1) To keep our algorithm computationally manageable, we relax the constraint on distributional consistency before and after collective anomalies in our algorithm. Although this may weaken detection power \citep{fisch2022subset, fisch2022linear}, the Bayesian framework mitigates the impact because it integrates over all possible segmentations. This allows imperfect segmentations to retain high marginal likelihoods, thereby increasing the posterior probability of collective anomalies; (2) Those brief transitional phases at change points, referred to as \emph{spurious anomalies}, can be misidentified as collective anomalies. To address this issue, we validate each collective anomaly by examining the presence of nearby change points. To restore the distributional consistency, we can remove collective anomalies upon detection. Alternatively, we could use interpolation, which might be more suitable for time series with temporal dependencies; and (3) Unlike our model, our algorithm allows two collective anomalies to occur consecutively.

\subsection{Calculating data likelihoods}
\label{subsec:algorithm1_loglik}

At each time $t$, we first compute $\{\mathbb{L}(\bm{y}^{t':t})\}$ and $\{\mathbb{P}(y^t|\bm{y}^{t':(t-1)})\}$ for time $t'\in \Upsilon_c^t$, defined as:
\begin{description}[noitemsep, topsep=2pt]
  \item[$\mathbb{L}(\bm{y}^{t':t})$] Conditional marginal likelihood of $\bm{y}^{t':t}$ given $t'=t-r^t$ and with $\theta^{t':t}$ integrated out, that is, $\mathrm{Pr}(\bm{y}^{t':t}|r^t=t-t')$;
  \item[$\mathbb{P}(y^t|\bm{y}^{t':(t-1)})$] Conditional probability of $y^t$ given $\bm{y}^{t':(t-1)}$ and $t'=t-r^t$, and with $\theta^{t':t}$ integrated out, that is, $\mathrm{Pr}(y^t|\bm{y}^{t':(t-1)},r^t=t-t')$.
\end{description}
The functional form of $\mathbb{L}(\bm{y}^{t':t})$ depends on data distribution $\mathrm{Pr}(\bm{y}^{t':t}|\theta^t)$ and prior $\mathrm{Pr}(\theta^{t':t}|\pi_0)$, and can be approximated via numerical integration when an analytical expression is unavailable \citep{fearnhead2006exact}. The posterior predictive probability $\mathbb{P}(y^t|\bm{y}^{t':(t-1)})$ can be calculated as $\mathbb{L}(\bm{y}^{t':t})/\mathbb{L}(\bm{y}^{t':(t-1)})$ based on the Bayes theorem.


Then, we recursively compute three types of data likelihoods to infer the most recent collective anomaly and change point:
\begin{description}[noitemsep, topsep=2pt]
  \item[$\mathbb{W}^t_a(d,r)$] Likelihood of $\bm{y}^{1:t}$ with the most recent change point at time $t-d$ and the most recent collective anomaly ending at time $t-r$ after the change point, that is,
  $\mathrm{Pr}(\bm{y}^{1:t},d^t=d,r^t=r,a^{t-r}=1)$ for $d\leq n_c^t$ and $r\geq t-R(t,d)+1$;
  \item[$\mathbb{W}^t_c(d)$] Likelihood of $\bm{y}^{1:t}$ with the most recent change point at time $t-d$ and no collective anomaly between time $t-d$ and $t$, that is,
  $\mathrm{Pr}(\bm{y}^{1:t},d^t=d,r^t=d)$ for $d\leq n_c^t$;
  \item[$\mathbb{Q}^t_c(d)$] Likelihood of $\bm{y}^{1:t}$ with the most recent change point at time $t-d$, that is,
  $\mathrm{Pr}(\bm{y}^{1:t},d^t=d)$ for $d\leq n_c^t$.
\end{description}
The above likelihoods will be used to derive the posterior probabilities and MAP estimates of $r^t$, $a^{t-r^t}$, and $d^t$ in Equations (\ref{equ:r_t2})-(\ref{equ:d_t_post}). A set of recursion functions for calculating these data likelihoods is given by the following theorem:
\begin{theorem}
\label{thm1}
Define likelihoods $\mathbb{W}^t_a\left(d,r\right)$, $\mathbb{W}^t_c\left(d\right)$, and $\mathbb{Q}^t_c\left(d\right)$ as above. The recursion deriving them is initialized at time $t=1$ with $\mathbb{W}^1_a\left(0,0\right)=0$, $\mathbb{W}^1_c\left(0\right)=\mathbb{L}\left(y^{1}\right)$, $\mathbb{Q}^1_c\left(0\right)=\mathbb{L}\left(y^{1}\right)$, and updated at time $t>1$ with:
\begin{equation}
    \mathbb{W}^t_a\left(d,r\right)=
    \left\{
    \begin{aligned}
    &\mathbb{W}^{t-1}_a\left(d-1,r-1\right) \mathbb{P}\left(y^t\left|\bm{y}^{\left(t-r\right):\left(t-1\right)}\right.\right) \left(1-p_0\right),\ \text{if}\ r>0,\\
    &\sum_{r'=0}^{A(t,d)} \mathbb{Q}^{t-2-r'}_c\left(d-2-r'\right)\mathbb{L}\left(\bm{y}^{\left(t-1-r'\right):\left(t-1\right)}\right)
    \mathbb{L}\left(y^{t}\right) p_0\left(1-q_0\right)^{r'}q_0 ,\ \text{otherwise},\\
    \end{aligned}
    \right.
    \label{equ:w_a}
\end{equation}
\begin{equation}
    \mathbb{W}^t_c\left(d\right)=
    \left\{
    \begin{aligned}
    &\mathbb{W}^{t-1}_c\left(d-1\right) \mathbb{P}\left(y^t\left|\bm{y}^{\left(t-d\right):\left(t-1\right)}\right.\right) \left(1-p_0\right),\ \text{if}\ d>\Delta t\ \text{or}\ d=t-1,\\
    &\mathbb{W}^{t-1}_c\left(d-1\right) \mathbb{P}\left(y^t\left|\bm{y}^{\left(t-d\right):\left(t-1\right)}\right.\right) \left(1-q_0\right),\ \text{if}\ 0<d\leq\Delta t\ \text{and}\ d\neq t-1,\\
    &\sum_{d'=\Delta t}^{n_c^{t-1}} \mathbb{Q}^{t-1}_c\left(d'\right)
    \mathbb{L}\left(y^{t}\right) p_0,\ \text{if}\ d=0\ \text{and}\ t\geq \Delta t+3,\\
    & \mathbb{Q}^{t-1}_c\left(t-2\right)
			\mathbb{L}\left(y^{t}\right) p_0,\  \text{otherwise},\\
    \end{aligned}
    \right.
    \label{equ:w_c}
\end{equation}
\begin{equation}
    \mathbb{Q}^t_c\left(d\right)=\sum_{r=0}^{R(t,d)-1}\mathbb{W}^t_a\left(d,r\right)+\mathbb{W}^t_c\left(d\right).
    \label{equ:q_c}
\end{equation}
\end{theorem}
The proof is provided in the supplementary material.

\subsection{Calculating posterior probabilities}
\label{subsec:algorithm1_posterior}

First, we perform collective anomaly detection while accounting for potential change points. At time $t$, the posterior probability of $r^t=r$ for $r\in\{0,\dots,n_c^t\}$ can be calculated via the Bayes theorem and the law of total probability:
\begin{equation}
    \mathrm{Pr}\left(r^t=r\left|\bm{y}^{1:t}\right.\right)=\frac{\sum_{d=r+2}^{n_c^t}\mathbb{W}^t_a\left(d,r\right)+\mathbb{W}^t_c\left(r\right)}{\sum_{r'=0}^{n_c^t}\left(\sum_{d=r'+2}^{n_c^t}\mathbb{W}^t_a\left(d,r'\right)+\mathbb{W}^t_c\left(r'\right)\right)}.
    \label{equ:r_t2}
\end{equation}
The MAP estimate of $r^t$, denoted as $r^*$, equals $\argmax_{0\leq r\leq n_c^t} \sum_{d=r+2}^{n_c^t}\mathbb{W}^t_a\left(d,r\right)+\mathbb{W}^t_c\left(r\right)$. Accordingly, $t-r^*$ is the MAP estimate of the timing of the most recent change.



If the most recent change at time $t-r^*$ falls within search time range $\Upsilon_a^t$, that is, if $r^*\leq n_a^t$, we examine whether it indicates a collective anomaly. Ideally, this examination should rely on the conditional posterior probability that the most recent anomaly occurs at time $t-r^*$, $\mathrm{Pr}(a^{t-r^t}=1, r^{t-r^t-1}\leq \Delta t-1, (t-r^t-1)-r^{t-r^t-1}\leq t-r^* |r^*-\Delta t\leq r^t\leq r^*,\bm{y}^{1:t})$. Since calculating this probability is non-trivial, we employ a tractable approach:
\begin{equation}
    \mathrm{Pr}\left(a^{t-r^t}=1\left|r^*-\Delta t\leq r^t\leq r^*,\bm{y}^{1:t}\right.\right)=
    \frac{\sum_{r=max\{0,r^*-\Delta t\}}^{r^*} \sum_{d=r+2}^{n_c^t}\mathbb{W}^t_a\left(d,r\right)}{\sum_{r=max\{0,r^*-\Delta t\}}^{r^*} \left(\sum_{d=r+2}^{n_c^t}\mathbb{W}^t_a\left(d,r\right)+\mathbb{W}^t_c\left(r\right)\right)}.
    \label{equ:a_t_post1}
\end{equation}
Since $\{\sum_{d=r+2}^{n_c^t}\mathbb{W}^t_a(d,r)\}$ was computed when obtaining $r^*$, recalculation is unnecessary here. We trigger an alert for an anomaly if the probability exceeds a pre-specified threshold, $\lambda_a$. We now explain why the loss of applying this probability is negligible. When an anomaly truly occurs, the detection power remains unaffected because this probability is always larger than the first one. Otherwise, the most recent change at time $t-r^*$ is a change point, and the difference between the two probabilities becomes small when the change is significant and $q_0$ is small. Moreover, even if our algorithm falsely recognizes an anomaly, subsequent examination will classify it as a spurious anomaly, thereby avoiding the false alarm.


Once a new anomaly is detected, we identify its start and end times, and classify it. First, we derive the MAP estimates for its end time $t-r_1$ and start time $(t-r_1-1)-r_2$ sequentially:
\begin{equation}
    r_1=\argmax_{r^*-\Delta t\leq r\leq  r^*}\sum_{d=r+2}^{n_c^t}\mathbb{W}^t_a\left(d,r\right),\quad
    r_2=\argmax_{0\leq r\leq \Delta t-1}\mathbb{W}^{t-r_1-1}_c\left(r\right),
    \nonumber
\end{equation}
where $\{\mathbb{W}^{t-r_1-1}_c(r)\}$ were derived and stored at time $t-r_1-1\in \Upsilon_a^t\cup\{t-n_a^t-1\}$. Then, we remove times $(t-r_1-1)-r_2,\dots,t-r_1-1$ from search time ranges $\Upsilon_a^t$ and $\Upsilon_c^t$, and recalculate the likelihoods affected by this removal. The above detection and removal procedure is repeated until no more anomalies are identified. In this way, we identify a series of recent anomalies rather than just the most recent one. After reestimating the most recent change at time $t-r^*$, a removed anomaly is classified as a collective anomaly if it is distant from the change, and as a spurious anomaly otherwise. If collective anomalies are retained in the time series in subsequent detection, we revert $\Upsilon_a^t$, $\Upsilon_c^t$, and the involved likelihoods to their previous values. Specifically, spurious anomalies must be removed to prevent BOCD from overlooking nearby change points.

Finally, we update the estimate for the most recent change point at time $t-d^t$. The posterior probability of $d^t=d$ for $d\in \{0,\dots,n_c^t\}$ can be calculated via the Bayes theorem:
\begin{equation}
    \mathrm{Pr}\left(d^t=d\left|\bm{y}^{1:t}\right.\right)=\frac{\mathbb{Q}^t_c\left(d\right)}{\sum_{d'=0}^{n_c^t}\mathbb{Q}^t_c\left(d'\right)}.
    \nonumber
\end{equation}
The MAP estimate of $d^t$ is $d^*=\argmax_{d}\mathbb{Q}^t_c(d)$, implying that the most recent change point occurs at time $t-d^*$.
To account for the uncertainty in localizing a change point in fine-grained time series, we compute the posterior probability of a change point within a time window of length $2\delta+1$ centered at time $t-d^*$:
\begin{equation}
    \mathrm{Pr}\left(d^*-\delta \leq d^t \leq d^*+\delta \left|\bm{y}^{1:t}\right.\right)=\frac{\sum_{d'=\max\{d^*-\delta,0\}}^{\min\{d^*+\delta,n_c^t\}}\mathbb{Q}^t_c\left(d'\right)}{\sum_{d'=0}^{n_c^t}\mathbb{Q}^t_c\left(d'\right)},
    \label{equ:d_t_post}
\end{equation}
where $\delta$ is the tolerance for uncertainty in change point localization. We trigger an alert for a change point when this posterior probability exceeds a pre-specified threshold, $\lambda_c$.

\section{Approximation algorithm with reduced time and space complexity}
\label{sec:algorithm2}

To reduce time and space complexity, we develop another Bayesian online change detection algorithm enhanced with collective anomaly removal (BOCD-AR). In this algorithm, collective anomalies are removed immediately upon detection, and change points are declared only after sufficient post-change data have been observed. When change classification is both accurate and timely, this strategy ensures that no detectable collective anomalies remain in the time series following a declared change point. Under this condition, the most recent change corresponds to a change point, allowing us to reliably approximate $d^t$ by $r^t$. In this section, we first detail the calculation of data likelihoods and posterior probabilities, and then conduct a complexity analysis. Both time and space complexity grow linearly with the length of the search time range. 




\subsection{Calculating data likelihoods}
\label{subsec:algorithm2_loglik}

At each time $t$, we first compute conditional marginal likelihoods $\{\mathbb{L}(\bm{y}^{t':t})\}$ and posterior predictive probabilities $\{\mathbb{P}(y^t|\bm{y}^{t':(t-1)})\}$ for $t'\in \Upsilon_c^t$, as in Section \ref{subsec:algorithm1_loglik}. Then, we recursively calculate two types of likelihoods to infer the most recent collective anomaly and change point: 
\begin{description}[noitemsep, topsep=2pt]
  \item[$\mathbb{H}^t_a(r)$] Likelihood of $\bm{y}^{1:t}$ with the most recent change at time $t-r$ with $a^{t-r}=1$, that is,
  $\mathrm{Pr}(\bm{y}^{1:t},r^t=r,a^{t-r}=1)$ for $r\leq n_c^t$;
  \item[$\mathbb{H}^t_c(r)$] Likelihood of $\bm{y}^{1:t}$ with the most recent change at time $t-r$ with $a^{t-r}=0$, that is,
  $\mathrm{Pr}(\bm{y}^{1:t},r^t=r,a^{t-r}=0)$ for $r\leq n_c^t$.
\end{description}
These data likelihoods can be recursively calculated by:
\begin{theorem}
	\label{thm2}
	Define likelihoods $\mathbb{H}^t_a(r)$ and $\mathbb{H}^t_c(r)$ as above. The recursion deriving them is initialized at time $1$ with $\mathbb{H}^1_a\left(0\right)=0$ and $ \mathbb{H}^1_c\left(0\right)=\mathbb{L}\left(y^{1}\right)$, and updated at time $t>1$ with:
	\begin{equation}
		\mathbb{H}^t_a\left(r\right)=
		\left\{
		\begin{aligned}
			&\mathbb{H}^{t-1}_a\left(r-1\right) \mathbb{P}\left(y^t\left|\bm{y}^{\left(t-r\right):\left(t-1\right)}\right.\right) \left(1-p_0\right)
			,\ \text{if}\ r> 0,\\
			&\sum_{r'=0}^{A(t,t-1)}\mathbb{H}^{t-1}_c\left(r'\right)\mathbb{L}\left(y^{t}\right) q_0,\ \text{otherwise},\\
		\end{aligned}
		\right.
		\label{equ:h_a}
	\end{equation}
	\begin{equation}
		\mathbb{H}^t_c\left(r\right)=
		\left\{
		\begin{aligned}
			&\mathbb{H}^{t-1}_c\left(r-1\right) \mathbb{P}\left(y^t\left|\bm{y}^{\left(t-r\right):\left(t-1\right)}\right.\right) \left(1-p_0\right)
			,\ \text{if}\ r>\Delta t\ \text{or}\ r= t-1,\\
			&\mathbb{H}^{t-1}_c\left(r-1\right) \mathbb{P}\left(y^t\left|\bm{y}^{\left(t-r\right):\left(t-1\right)}\right.\right) \left(1-q_0\right)
			,\ \text{if}\ 0<r\leq \Delta t\ \text{and}\ r\neq t-1,\\
			&\left(
			\sum_{r'=\Delta t}^{n_c^{t-1}}\mathbb{H}^{t-1}_c\left(r'\right)
			+\sum_{r'=0}^{n_c^{t-1}}\mathbb{H}^{t-1}_a\left(r'\right)
			\right)\mathbb{L}\left(y^{t}\right)p_0
			,\ \text{if}\ r=0\ \text{and}\ t\geq\Delta t+3
            \\
            &\left(\mathbb{H}^{t-1}_c\left(t-2\right)
			+\sum_{r'=0}^{n_c^{t-1}}\mathbb{H}^{t-1}_a\left(r'\right)
			\right)\mathbb{L}\left(y^{t}\right)p_0,\ \text{otherwise}.\\
		\end{aligned}
		\right.
		\label{equ:h_c}
	\end{equation}
\end{theorem}

The end and start times of the most recent collective anomaly can be sequentially identified using likelihoods $\{\mathbb{H}^t_a(r)\}$ and $\{\mathbb{H}^{t'}_c(r)\}$, as in BOCD. Alternatively, we could perform joint estimation with the third type of likelihoods, $\{\mathbb{G}^t_a(r',r)\}$, where $\mathbb{G}^t_a(r',r)$ denotes the likelihood of $\bm{y}^{1:t}$ with the most recent two changes at time $t-r\in \Upsilon_a^t$ and time $(t-r-1)-r'\in\{(t-r-1)-A(t-r,t-r-1),\dots,t-r-1\}$ enclosing a collective anomaly. These likelihoods can be recursively computed by:
\begin{theorem}
	\label{thm3}
	Define likelihood $\mathbb{G}^t_a(r',r)$ as above. The recursion for calculating $\mathbb{G}^t_a(r',r)$ is initialized at time $t=3$ with $\mathbb{G}^3_a\left(0,0\right)=\mathbb{L}\left(y^{1}\right)\mathbb{L}\left(y^{2}\right)\mathbb{L}\left(y^{3}\right)p_0q_0$, and updated at time $t>3$ with:
	\begin{equation}
		\mathbb{G}^t_a\left(r',r\right)=
		\left\{
		\begin{aligned}
			&\mathbb{G}^{t-1}_a\left(r',r-1\right)\mathbb{P}\left(y^t\left|\bm{y}^{\left(t-r\right):\left(t-1\right)}\right.\right)\left(1-p_0\right),\ \text{if}\ r>0,\\
			&\mathbb{H}^{t-1}_c\left(r'\right)\mathbb{L}\left(y^{t}\right) q_0,\ \text{otherwise},\\
		\end{aligned}
		\right.
		\label{equ:g_a}
	\end{equation}
	where data likelihood $\mathbb{H}^{t-1}_c(r')$, as defined previously, is derived from Theorem \ref{thm2}.
\end{theorem}
The proofs of Theorems \ref{thm2} and \ref{thm3} are provided in the supplementary material.

\subsection{Calculating posterior probabilities}
\label{subsec:algorithm2_posterior}

The posterior probabilities used for change detection are similar to those in BOCD. At time $t$, we first examine the existence of a collective anomaly near the most recent change. The posterior probability of $r^t=r$ for $r\in \{0,\dots,n_c^t\}$ can be calculated via the Bayes theorem:
\begin{equation}
	\mathrm{Pr}\left(r^t=r\left|\bm{y}^{1:t}\right.\right)=\frac{\mathbb{H}^{t}_a\left(r\right)+\mathbb{H}^{t}_c\left(r\right)}{\sum_{r'=0}^{n_c^t}\left(\mathbb{H}^{t}_a\left(r'\right)+\mathbb{H}^{t}_c\left(r'\right)\right)}.
    \label{equ:r_t_post}
\end{equation}
We identify the MAP estimate of $r^t$ as $r^*=\argmax \mathbb{H}^{t}_a\left(r\right)+\mathbb{H}^{t}_c\left(r\right)$. If the most recent change at time $t-r^*$ falls within search time range $\Upsilon_a^t$, we further examine the presence of a collective anomaly nearby based on the conditional posterior probability:
\begin{equation}
	\begin{aligned}
		&\mathrm{Pr}\left(a^{t-r^t}=1, 
		r^{t-r^t-1}\leq \Delta t-1,
		\left(t-r^t-1\right)-r^{t-r^t-1}\leq t-r^*
		\left|r^*-\Delta t\leq r^t\leq r^*,\bm{y}^{1:t}\right.\right)\\
		=&\frac{\sum_{r=max\{0,r^*-\Delta t\}}^{r^*}
			\sum_{r'=r^*-r-1}^{A(t-r,t-r-1)}
			\mathbb{G}^t_a\left(r',r\right)}
		{\sum_{r=max\{0,r^*-\Delta t\}}^{r^*}\left(\mathbb{H}^{t}_a\left(r\right)+\mathbb{H}^{t}_c\left(r\right)\right)}.\\
	\end{aligned}
	\label{equ:a_t_post}
\end{equation}
Computing the numerator requires a time complexity of $\mathcal{O}(\Delta t^2)$, which can be substantial when $\Delta t$ is large. Similar to BOCD, we can use the probability that does not involve $\{\mathbb{G}^t_a(r',r)\}$ instead:
\begin{equation}
	\mathrm{Pr}\left(a^{t-r^t}=1\left|r^*-\Delta t\leq r^t\leq r^*,\bm{y}^{1:t}\right.\right)=
	\frac{\sum_{r=max\{0,r^*-\Delta t\}}^{r^*} \mathbb{H}^{t}_a\left(r\right)}{\sum_{r=max\{0,r^*-\Delta t\}}^{r^*}\left(\mathbb{H}^{t}_a\left(r\right)+\mathbb{H}^{t}_c\left(r\right)\right)},
	\label{equ:a_t_post2}
\end{equation}
which reduces the complexity to $\mathcal{O}(\Delta t)$. We trigger an alert for a collective anomaly when the probability exceeds threshold $\lambda_a$. 

Once a new collective anomaly is detected, we identify its start and end times, and remove it from search time ranges $\Upsilon_a^t$ and $\Upsilon_c^t$. The end and start times, denoted as $t-r_1$ and $(t-r_1-1)-r_2$, can be inferred either jointly or sequentially. For joint inference, we derive the MAP estimate of $(r_2,r_1)$ as $\argmax_{r_2,r_1}\mathbb{G}^t_a(r_2,r_1)$. For sequential inference, we first identify the MAP estimate of $r_1$ as $\argmax_{r_1}\mathbb{H}^{t}_a(r_1)$, and then that of $r_2$ as $\argmax_{r_2}\mathbb{H}^{t-r_1-1}_c(r_2)$. This requires retaining the values of $\mathbb{H}^{t'}_c(r')$ for $r'=0,\dots,\Delta t-1$ computed at time $t'\in \Upsilon_a^t\cup\{t-n_a^t-1\}$. The anomaly detection and removal procedure is repeated until no more anomalies are identified. Each detected collective anomaly is validated by its distance to the updated most recent change.


Finally, we infer the most recent change point at time $t-r^t$. The MAP estimate $r^*$ was derived from Equation (\ref{equ:r_t_post}). As with BOCD, we calculate the posterior probability that the most recent change point falls within a time window of length $2\delta+1$ centered at time $t-r^*$:
\begin{equation}
    \mathrm{Pr}\left(r^*-\delta\leq r^t \leq r^*+\delta\}\left|\bm{y}^{1:t}\right.\right)=\frac{\sum_{r'=\max\{r^*-\delta,0\}}^{\min\{r^*+\delta,n_c^t\}}\left(\mathbb{H}^{t}_a\left(r'\right)+\mathbb{H}^{t}_c\left(r'\right)\right)}{\sum_{r'=0}^{n_c^t}\left(\mathbb{H}^{t}_a\left(r'\right)+\mathbb{H}^{t}_c\left(r'\right)\right)}.
    \label{equ:r_t_post2}
\end{equation}
We consider the change point at time $t-r^*$ significant if this probability exceeds $\lambda_c$.

\subsection{Complexity analysis}
\label{subsec:complexity_algo2}

We now discuss the time complexity at time $t$ based on the overall detection procedure described in Algorithm \ref{alg:detect2}. In likelihood calculation, the first loop operates within $\mathcal{O}(u_c)$ time bound. In the second loop, Equation (\ref{equ:h_a}) takes $\mathcal{O}(1)$ time when $r>0$ and $\mathcal{O}(\Delta t)$ time when $r=0$. Meanwhile, Equation (\ref{equ:h_c}) takes $\mathcal{O}(1)$ time when $r>0$ and $\mathcal{O}(u_c)$ time when $r=0$. Based on the above analysis, the second loop takes $\mathcal{O}(\Delta t+u_c)$ time. Since $u_c$ is typically larger than $\Delta t$, this complexity is equivalent to $\mathcal{O}(u_c)$. The third loop runs in $\mathcal{O}(u_a\Delta t)$ time bound as Equation (\ref{equ:g_a}) takes $\mathcal{O}(1)$ time. Next, the identification of $r^*$ and the computation of Equation (\ref{equ:a_t_post2}) take $\mathcal{O}(u_c)$ and $\mathcal{O}(\Delta t)$ time, respectively. 
Given that no new collective anomaly is detected, the complexity of collective anomaly detection is $\mathcal{O}(u_c)$. Specifically, if we compute the probability in Equation (\ref{equ:a_t_post}) instead of that in Equation (\ref{equ:a_t_post2}), the complexity increases to $\mathcal{O}(u_c+\Delta t^2)$. The process of change point detection takes $\mathcal{O}(u_c)$ time. Based on the above analysis, the overall complexity of our new algorithm is $\mathcal{O}(u_c+u_a\Delta t)$ when no new anomaly is detected. Furthermore, the calculation of $\{\mathbb{G}^t_a(r',r)\}$ can be avoided by employing an alternative method described in the second and third paragraphs of Section \ref{subsec:algorithm2_posterior}. This reduces the computational complexity to $\mathcal{O}(u_c)$, which is the same as that of BOCPD \citep{adams2007bayesian}. The computational complexity of BOCD-AR is substantially smaller than that of BOCD, $\mathcal{O}(u_c^2)$. This is because BOCD traverses all possible values of $(d^t,r^t)$ to calculate the corresponding likelihoods, while BOCD-AR only traverses the possible values of $r^t$.

\begin{algorithm}[!h]
	\caption{Overall detection procedure at time $t$.}
	\label{alg:detect2}
	\KwIn{Hyperparameters $\{p_0, q_0, \Delta t, \lambda_a, \lambda_c, \delta\}$, search time ranges $\Upsilon_a^t$ and $\Upsilon_c^t$, time series data $\bm{y}^{(t-n_c^t):t}$, likelihoods $\{\mathbb{L}(\bm{y}^{t'':t')})\}$, $\{\mathbb{H}^{t'}_a(r)\}$, $\{\mathbb{H}^{t'}_c(r)\}$, and (optional) $\{\mathbb{G}^{t'}_a(r',r)\}$ for $t'<t$;}
	\KwOut{The most recent change point at time $t-r^*$, start and end times of collective anomalies $\{((t-r_1-1)-r_2,t-r_1)\}$, updated search time ranges and likelihoods;}
	\For{$t'\in\Upsilon_c^t$}{
		Calculate $\mathbb{L}(\bm{y}^{t':t})$ and compute $\mathbb{P}(y^t|\bm{y}^{t':(t-1)})\leftarrow \mathbb{L}(\bm{y}^{t':t})/\mathbb{L}(\bm{y}^{t':(t-1)})$\;
	}
	\For{$r=0,\dots,n_c^t$}{
		Calculate $\mathbb{H}^t_a(r)$ and $\mathbb{H}^t_c(r)$ based on Equations (9) and (10)\;
	}
	\For{$r=0,\dots,n_a^t$}{
		\For{$r'=0,\dots,A(t-r,t-r-1)$}{
			(Optional) Calculate $\mathbb{G}^t_a(r',r)$ based on Equation (\ref{equ:g_a})\;
		}
	}
        Identify $r^*\in\{0,\dots, n_c^t\}$ that maximizes $\mathbb{H}^{t}_a(t-r)+\mathbb{H}^{t}_c(t-r)$\;
        Detect and remove collective anomalies based on the process described in the supplementary material\;
	Calculate $\mathrm{Pr}(r^*-\delta\leq r^t\leq r^*+\delta|\bm{y}^{1:t})$ based on Equation (\ref{equ:r_t_post2})\;
	\If{$\mathrm{Pr}(r^*-\delta\leq r^t\leq r^*+\delta|\bm{y}^{1:t})>\lambda_c$}{
		Alarm the updated most recent change point\;
	}
\end{algorithm}

Finally, let us analyze the space complexity at time $t$. We store likelihoods $\{\mathbb{L}(\bm{y}^{t':t})\}$, $\{\mathbb{H}^t_a(r)\}$, and $\{\mathbb{H}^t_c(r)\}$ for the recursion at time $t+1$, which all require $\mathcal{O}(u_c)$ space. Besides, the storage of $\{\mathbb{G}^t_a(r',r)\}$ for the recursion at time $t+1$ takes $\mathcal{O}(u_a\Delta t)$ space. If we eliminate $\{\mathbb{G}^t_a(r',r)\}$ using the alternative method, we store the values of $\mathbb{H}^{t'}_c(r)$ for $r=0,\dots,\Delta t-1$ at time $t'\in \Upsilon_a^t \cup\{t-n_a^t-1\}$, which requires the same $\mathcal{O}(u_a\Delta t)$ space. To recalculate likelihoods after anomaly removal, we need to retrieve $\{\mathbb{L}(\bm{y}^{t'':t'})\}$, $\{\mathbb{H}^{t'}_a(r)\}$, and $\{\mathbb{H}^{t'}_c(r)\}$ for $u_a+\Delta t$ possible values of $t'=(t-r_1-1)-r_2-1$, or, alternatively, for $t'=t-n_a^t-\Delta t-1$ with additional computation cost. The former requires $\mathcal{O}(u_c(u_a+\Delta t))$ space, and the latter requires $\mathcal{O}(u_c)$ space. When $\{\mathbb{G}^{t'}_a(r',r)\}$ are also retrieved, the space complexity becomes $\mathcal{O}(u_a\Delta t(u_a+\Delta t))$ for the former case and $\mathcal{O}(u_a\Delta t)$ for the latter. Based on the above analysis, the minimal space complexity of our algorithm is $\mathcal{O}(u_c+u_a\Delta t)$. Notably, this is smaller than that of BOCD, $\mathcal{O}(u_c^2)$, and comparable to that of BOCPD \citep{adams2007bayesian}, $\mathcal{O}(u_c)$.

\section{Choices for hyperparameters}
\label{sec:hyperparameter}

We discuss appropriate choices for our hyperparameters, including prior change probabilities $p_0$ and $q_0$, the maximum duration of a collective anomaly $\Delta t$, the upper limits for the lengths of search time ranges $u_a+1$ and $u_c+1$, detection thresholds $\lambda_a$ and $\lambda_c$, and uncertainty tolerance $\delta$. A small $p_0$ is commonly used due to the sparsity of changes. The choices of $\Delta t$ and $\delta$ are determined by the temporal granularity of the data. Thresholds $\lambda_a$ and $\lambda_c$ are chosen based on the application context and the trade-off between false positives and false negatives. A reasonably large $u_a$ is required to ensure that $u_a$ exceeds the expected delay in detecting a collective anomaly. Similarly, $u_c$ should be large enough for $\Upsilon_c^t$ to contain the most recent change point. Alternatively, $u_c$ can be set by truncating the tail of the posterior distribution of $r^t$ where the cumulative probability mass falls below a small threshold \citep{adams2007bayesian}.

\begin{figure}[h]
	\centering
	\begin{subfigure}[b]{0.46\textwidth}
		\includegraphics[width=\textwidth]{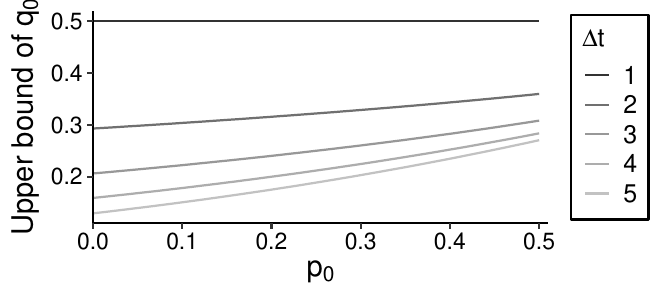}
        \caption{}
        \label{fig:hyper_bound_q_0}
	\end{subfigure}
	\begin{subfigure}[b]{0.46\textwidth}
		\includegraphics[width=\textwidth]{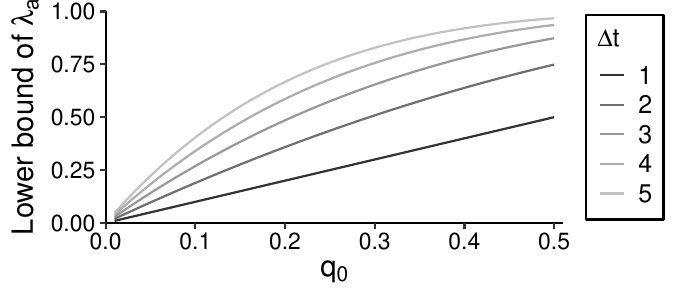}
        \caption{}
        \label{fig:hyper_bound_lambda_a}
	\end{subfigure}
	\caption{The relations (a) between $p_0$ and the upper bound of $q_0$ under $\lambda_a=0.5$, and (b) between $q_0$ and the lower bound of $\lambda_a$ under $p_0=0.01$.}
	\label{fig:hyper_bound}
\end{figure}

Here, we focus on $q_0$ and $\lambda_a$, which are the prior for change classification and the threshold for declaring an anomaly, respectively. A large $q_0$ helps to detect less significant collective anomalies. This, however, can cause the probabilities in Equations (\ref{equ:a_t_post1}), (\ref{equ:a_t_post}), and (\ref{equ:a_t_post2}) to consistently exceed $\lambda_a$, leading to frequent spurious anomaly alarms and extra verification. To avoid this issue, $q_0$ and $\lambda_a$ can be chosen based on the following inequality, as explained in the supplementary material:
\begin{equation}
    \frac{p_0q_0\sum_{i=0}^{\Delta t-1} \left(1-q_0\right)^i\left(1-p_0\right)^{\Delta t-1-i}}{p_0q_0\sum_{i=0}^{\Delta t-1} \left(1-q_0\right)^i\left(1-p_0\right)^{\Delta t-1-i}+p_0\left(1-q_0\right)^{\Delta t}}<\lambda_a.
    \label{equ:hyper}
\end{equation}
An analytic solution for $q_0$ is not available. However, as proven in the supplementary material, the left side of this inequality increases monotonically with respect to $q_0$. Hence, we could identify the upper bound of $q_0$ through a line search. Figure \ref{fig:hyper_bound} presents the relation between $p_0$ and the upper bound of $q_0$ under $\lambda_a=0.5$, and that between $q_0$ and the lower bound of $\lambda_a$ under $p_0=0.01$.

\section{Simulation study}
\label{sec:simulation}

In this section, we validate the proposed method using simulated time series. We first describe the experimental setup. Then, we conduct a performance comparison, followed by a sensitivity analysis of hyperparameters. We visualize the online change detection procedure in the supplementary material. Our algorithms are implemented in Python. 

\subsection{Experimental setup}
\label{subsec:simu_setup}

\begin{figure}[h]
    \centering
    \includegraphics[width=\textwidth]{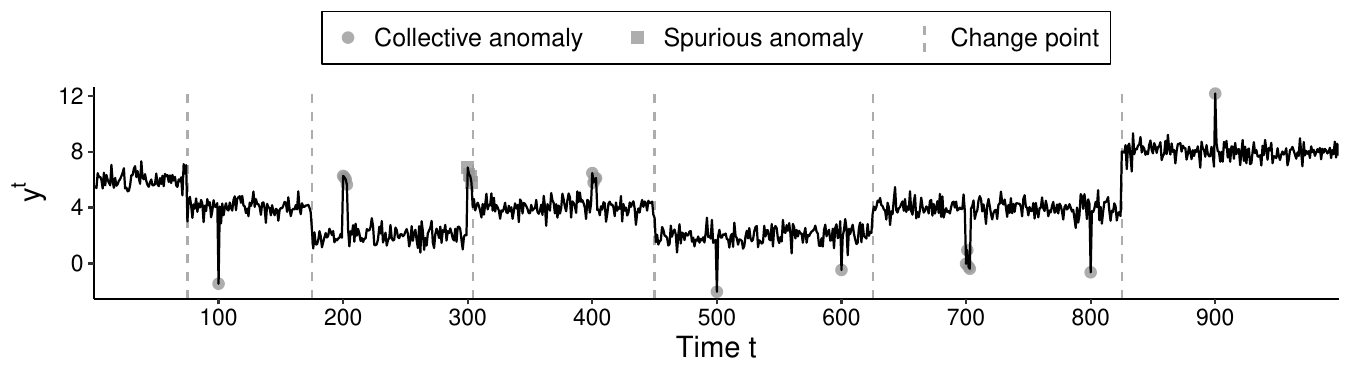}
    \caption{Simulated time series with anomalies and change points.}
    \label{fig:ex_simu_data}
\end{figure}

We generate our data based on the simulation design of \cite{gupta2022real}. We sample 1000 time series of length 1000 from normal distributions with shifting means and standard deviation 0.5. In each time series, we set 6 change points at times $\{75, 175, 300, 450, 625, 825\}$. The mean of each data segment is randomly drawn from $\{2,4,6,8\}$ under the constraint that adjacent segments are assigned different means. We add an anomaly every 100 time points. For each anomaly, its duration is randomly set to 1 or 4 time points, and its mean shift is sampled from $\{\pm 2,\pm 4\}$. All anomalies are collective anomalies, except for a spurious one at time 300. Figure \ref{fig:ex_simu_data} presents an example of our simulated time series. We apply intercept-only Bayesian linear regression (see the supplementary material) to model each data segment. 

The hyperparameter settings of our method are detailed below. We set the upper limits for the lengths of our search time ranges as $u_c+1=300$ and $u_a+1=28$. The maximum duration of a collective anomaly, $\Delta t$, is set to its true value, 4. We choose $p_0=0.1$ and $q_0=0.2$ by referring to Figure \ref{fig:hyper_bound_q_0}. We set detection thresholds $\lambda_a$ and $\lambda_c$ to 0.5. Since change points in the simulation data occur instantaneously, uncertainty tolerance $\delta$ is set to $0$. Change alerts are triggered after observing at least five post-change data points. Both spurious anomalies and collective anomalies are removed upon detection. The hyperparameters in Bayesian linear regression are set to $\sigma_0^2=0.25$, $v_0=1$, and $k_0=0.01$. 

\subsection{Performance evaluation}
\label{subsec:comparison}

We evaluate online change detection performance based on five criteria: precision, recall, F1 score, false positive rate, and detection delay. Precision is the proportion of correctly detected changes among all changes identified. Recall is the proportion of correctly detected changes among true changes. F1 score is the harmonic mean of precision and recall. This criterion is suitable for imbalanced classification tasks like change detection, which can be viewed as imbalanced classification of time points. False positive rate is the proportion of changes of one type misidentified as the other. Detection delay is the average time lag for changes to be accurately identified. Preferable methods exhibit higher precision, recall, and F1 score, along with lower false positive rate and detection delay. The formulas of these criteria are detailed in the supplementary material. 

First, we compare our approach with the baseline methods listed in Table \ref{tab:baseline_method}, and report the results in Tables \ref{tab:compare_baseline_an} and \ref{tab:compare_baseline_cp}. The best method for each criterion is labeled in bold. The last row reports the improvement of our method over the best baseline method. As indicated in Table \ref{tab:compare_baseline_an}, our method achieves more than a 35\% improvement in precision, F1 score, and detection delay for online collective anomaly detection. The inferior performance of baseline methods is mainly attributed to misclassification, limited sensitivity, and estimation bias caused by change points, which are effectively addressed by our method. Meanwhile, our method maintains a high recall and a low false positive rate. Regarding change point detection performance, Table \ref{tab:compare_baseline_cp} shows that our method outperforms the baselines by over 10\% in precision, F1 score, false positive rate, and detection delay. Compared to BOCPD, which is highly sensitive to change, our method significantly improves robustness to anomalies with only a modest loss in recall.

\begin{table}[!h]
	\small
	\centering
	\caption{Baseline anomaly and change point detection methods.}
	\label{tab:baseline_method}
	\begin{tabular}{cllcc}
		\toprule
		\rule{0pt}{10pt}
		\multirow{2}{*}{Abbreviation}&\multirow{2}{*}{Method}&\multirow{2}{*}{Reference}&\multicolumn{2}{c}{Scope}\\
		\cmidrule(lr){4-5}
		&&&A&C\\
		\midrule
		GOT&Grubbs outlier test&\cite{grubbs1949sample}&\checkmark&\\
		COT&Chisquared outlier test&\cite{dixon1950analysis}&\checkmark&\\
		VDT&Variant of Dixon test&\cite{dixon1950analysis}&\checkmark&\\
		TOD&Tukey outlier detection&\cite{tukey1977exploratory}&\checkmark&\\
		GLR&Generalized likelihood ratio test&\cite{hawkins2005statistical}&&\checkmark\\
		BOCPD&Bayesian online change point detection&\cite{adams2007bayesian}&&\checkmark\\
		RWBS&Robust wild binary segmentation&\cite{fearnhead2019changepoint}&&\checkmark\\
		ARC&Adversarially robust change point detection&\cite{li2021adversarially}&&\checkmark\\
		JEPO&Joint estimation of parameters and outliers&\cite{chen1993joint}&\checkmark&\checkmark\\
		BARD&Bayesian abnormal region detector&\cite{bardwell2017bayesian}&\checkmark&\checkmark\\
		CAPA&The collective and point anomalies&\cite{fisch2022linear}&\checkmark&\checkmark\\
		\bottomrule
	\end{tabular}
	\\\raggedright\small\textit{Notes.} For the scope of detection, ``A'' and ``C'' denote anomalies and change points, respectively.
\end{table}

\begin{table}[!h]
\small
	\centering
	\caption{Performance comparison criteria for online collective anomaly detection.}
	\label{tab:compare_baseline_an}
	\begin{tabular}{cccccc}
		\toprule
		\rule{0pt}{10pt}
		Method&Precision&Recall&F1 score&False positive rate&Detection delay\\
\midrule
GOT&0.551&0.640&0.592&0.138&3.677\\
COT&0.500&0.814&0.619&0.020&4.650\\
VDT&0.692&0.435&0.534&0.038&2.321\\
TOD&0.674&0.624&0.648&0.200&3.474\\
JEPO&0.596&0.688&0.638&0.021&3.676\\
BARD&0.630&0.262&0.370&0.070&5.688\\
CAPA&0.599&0.751&0.667&\textbf{0.007}&1.697\\
BOCD&0.941&0.860&0.899&0.009&0.037\\
BOCD-AR&\textbf{0.947}&\textbf{0.861}&\textbf{0.902}&0.008&\textbf{0.018}\\
Improvement&36.8\%&5.8\%&35.2\%&-14.3\%&98.9\%\\
		\bottomrule
	\end{tabular}
	\\\raggedright\small\textit{Notes.} For each criterion, the best method is labeled in bold. The last row reports the improvement of our method over the best baseline method.
\end{table}

\begin{table}[!h]
\small
	\centering
	\caption{Performance comparison criteria for online change point detection.}
	\label{tab:compare_baseline_cp}
	\begin{tabular}{cccccc}
		\toprule
		\rule{0pt}{10pt}
        Method&Precision&Recall&F1 score&False positive rate&Detection delay\\
\midrule
GLR&0.182&0.377 &0.245&0.141&1.903\\
BOCPD&0.289&\textbf{0.973}&0.446&0.687&0.574\\
RWBS&0.690&0.744&0.716&0.079&10.597\\
ARC&0.275&0.018&0.035&0.169&5.918\\
JEPO&0.848&0.877&0.862&0.075&5.329\\
BARD&0.538&0.707&0.611&0.288&0.956\\
CAPA&0.501&0.940&0.654&0.115&4.962\\
BOCD&\textbf{0.942}&0.968&\textbf{0.955}&\textbf{0.013}&0.840\\
BOCD-AR&0.928&0.971&0.949&0.022&\textbf{0.384}\\
Improvement&11.1\%&-0.2\%&10.8\%&82.7\%&33.1\%\\
		\bottomrule
	\end{tabular}
	\\\raggedright\small\textit{Notes.} For each criterion, the best method is labeled in bold. The last row reports the improvement of our method over the best baseline method.
\end{table}

\begin{figure}[h]
    \centering
    \includegraphics[width=0.9\textwidth]{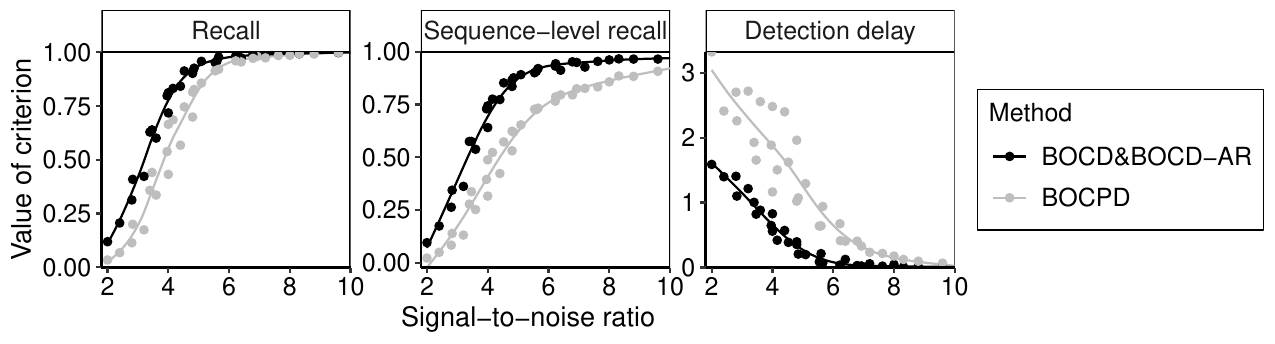}
    \caption{Performance comparison criteria across different SNR levels of a collective anomaly. The curves are generated using smoothing splines \citep{hastie2017generalized}. Sequence-level recall is the proportion of time series in which all detected changes coincide with the true collective anomalies.}
    \label{fig:minspace_jump}
\end{figure}

As collective anomalies can be viewed as short abnormal segments enclosed by change points, we further examine whether our method captures low-signal-to-noise-ratio (SNR) segments \citep{wang2020univariate} overlooked by change point approaches through an additional experiment. The experimental setup is detailed in the supplementary material. BOCD and BOCD-AR show nearly identical performance, and we compare them with BOCPD, which is highly sensitive to changes (see Table \ref{tab:compare_baseline_cp}). As shown in Figure \ref{fig:minspace_jump}, when the SNR is below 5.1, our method achieves an 8.5\% to 28.6\% higher recall. Moreover, our method requires, on average, 0.69 units less SNR to reach the same recall, as measured by the horizontal distance between the smoothing spline fits. When SNR exceeds 5.1, recall approaches 1 and further improvement becomes negligible. In terms of sequence-level recall, our method consistently outperforms BOCPD regardless of SNR. Moreover, our method significantly reduces detection delay, which is beneficial in low-SNR conditions where delays are typically longer.

\subsection{Comparison between the proposed algorithms}
\label{subsec:compare_proposed}

We further compare the detection performance of BOCD and BOCD-AR. As reported in the supplementary material, their recalls differ by less than 1.3\% and detection delays by less than 0.2 time points across different magnitudes and durations of change. Therefore, we focus on the changes where BOCD and BOCD-AR yield different detection results. A total of 105 data segments are identified as collective anomalies by only one of these algorithms, and most of them are misidentified by BOCD. Notably, about half result from BOCD misinterpreting their start times as change points and the short segments after their end times as collective anomalies. There are 90 change points detected by only one of our algorithms, most being collective anomalies misclassified as change points. Specifically, BOCD tends to model each of these anomalies with zero or one change point, while BOCD-AR models each with two. This is because BOCD tends to treat the end of an anomaly as the start of an undetected anomaly, instead of a change point.

Next, we compare the computational efficiency of BOCD and BOCD-AR. The experiment is conducted on a desktop computer with an Intel 2.00 GHz processor and 16GB of memory. We perform the experiment on a time series where both algorithms correctly detect all changes. Each algorithm is run 100 times to measure its average computation time. Results show that BOCD uses 36.671 s on average, while BOCD-AR takes only 10.844 s. As reported in Table \ref{tab:compute_time}, BOCD-AR is more efficient than BOCD across all detection steps, especially in the likelihood calculation. 

\begin{table}[!h]
\small
	\centering
	\caption{Average computation times for different steps of our algorithms.}
	\label{tab:compute_time}
	\begin{tabular}{lp{2cm}<{\centering}p{2cm}<{\centering}}
		\toprule
		\rule{0pt}{10pt}
		&\multicolumn{2}{c}{Average computation time (s)}\\
		\cmidrule(lr){2-3}
		&BOCD&BOCD-AR\\
		\midrule
		Likelihood calculation (excluding $\{\mathbb{L}(\bm{y}^{t':t})\}$)&9.329&0.300\\
		Detecting the most recent change&1.993&0.425\\
		Collective anomaly detection&0.012&0.004\\
		Detecting the most recent change point&0.012&--\\
		\bottomrule
	\end{tabular}
    \\\raggedright\small\textit{Notes.} We exclude the calculation of $\{\mathbb{L}(\bm{y}^{t':t})\}$ as it depends on the data distribution and the prior distribution. We also exclude the recalculation after anomaly removal. 
\end{table}

\subsection{Sensitivity analysis of hyperparameters}
\label{subsec:sensitivity}

\begin{figure}[h]
	\centering
	\includegraphics[width=\textwidth]{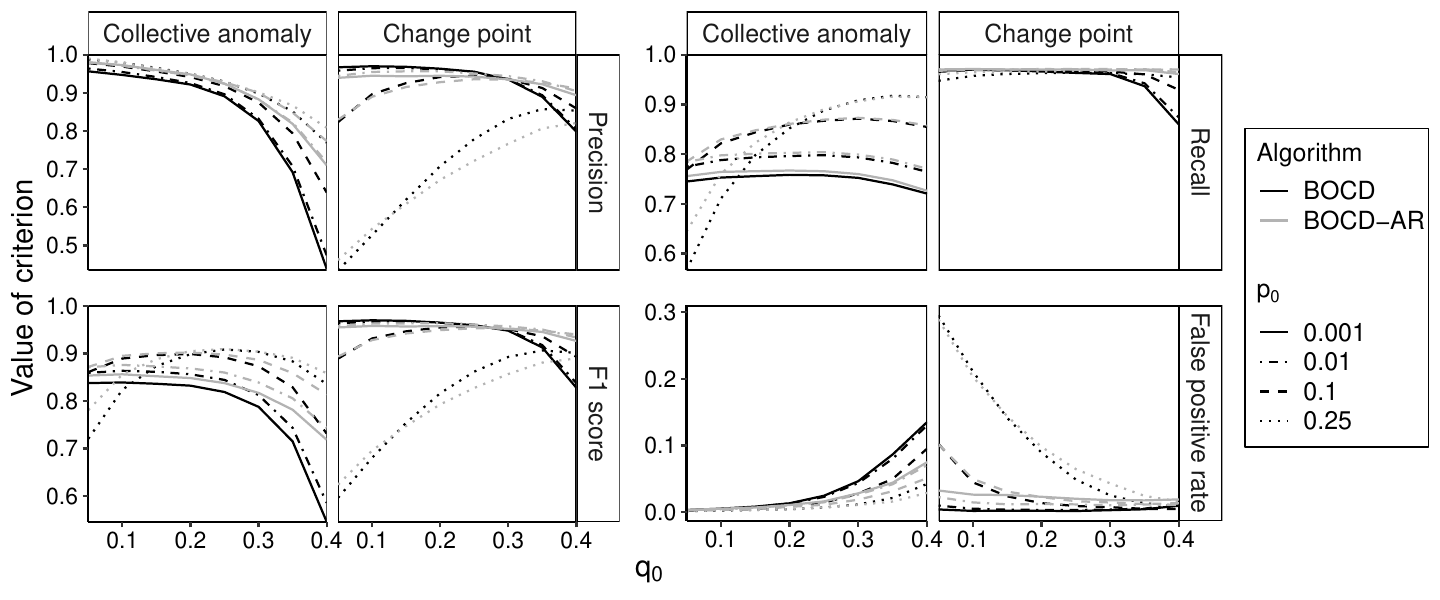}
	\caption{Detection accuracy criteria with different values of $p_0$ and $q_0$.}
	\label{fig:sensitivity_q0_method2}
\end{figure}

First, we assess how change detection accuracy varies with hyperparameters $q_0\in\{0.05,0.10,\dots,0.40\}$ and $p_0\in\{0.001, 0.01, 0.1, 0.25\}$ based on Figure \ref{fig:sensitivity_q0_method2}. The differences between the black and grey lines indicate that BOCD-AR is more robust to variations in $q_0$. Since these differences are small, we choose BOCD-AR for further analysis. The solid and dot-dash curves for $p_0=0.001$ and $p_0=0.01$ are rather similar, except that $p_0=0.01$ yields higher recall and F1 score in collective anomaly detection. Except recall, their criteria for collective anomaly detection degrade as $q_0$ increases, with a faster degradation rate at larger $q_0$. The dashed and dotted curves for $p_0=0.1$ and $p_0=0.25$ differ substantially from the curves for $p_0=0.001$ and $p_0=0.01$, and the former prefer a larger value of $q_0$. Overall, $p_0=0.1$ and $q_0\in[0.2,0.3]$ yield higher detection accuracy, although both values exceed the true change frequencies in the data. 

\begin{figure}[h]
	\centering
	\begin{subfigure}[b]{0.44\textwidth}
		\includegraphics[width=\textwidth]{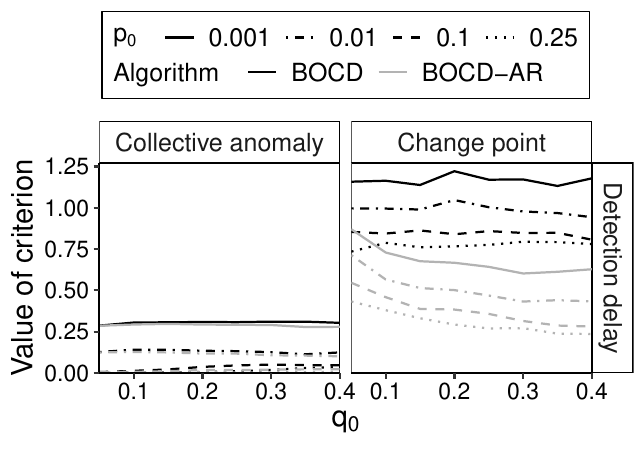}
        \caption{}
        \label{fig:ADD_q0_method2}
	\end{subfigure}
	\begin{subfigure}[b]{0.44\textwidth}
		\includegraphics[width=\textwidth]{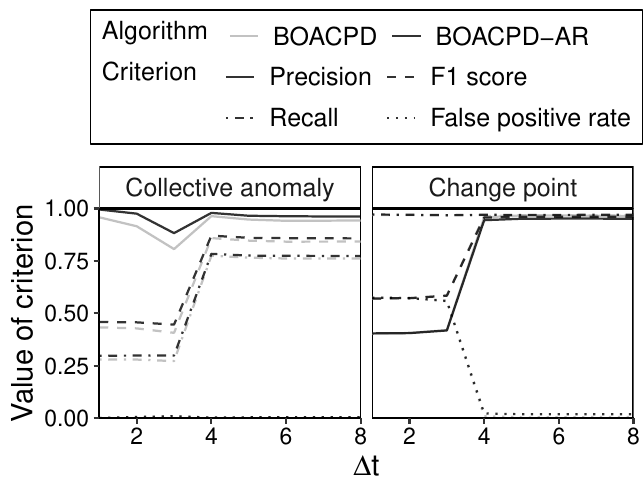}
        \caption{}
        \label{fig:sensitivity_delta_t_method2}
	\end{subfigure}
	\caption{(a) Detection delays with different values of $p_0$ and $q_0$. (b) Detection accuracy criteria with different values of $\Delta t$.}
	\label{fig:sensitivity_figure2}
\end{figure}

Then, we examine the impacts of $p_0$ and $q_0$ on detection delay based on Figure \ref{fig:ADD_q0_method2}. While BOCD and BOCD-AR have similar delays in detecting collective anomalies, BOCD-AR identifies change points more promptly. Besides, detection delay decreases with $p_0$ but is not sensitive to $q_0$. Therefore, among the hyperparameter settings that yield acceptable detection accuracy, a larger $p_0$ leads to a shorter detection delay and is recommended.

Finally, we explore the sensitivity of change detection accuracy to hyperparameter $\Delta t\in\{1,\dots,8\}$ based on Figure \ref{fig:sensitivity_delta_t_method2}. The curves for BOCD and BOCD-AR largely overlap. Detection accuracy declines significantly when $\Delta t$ is set below its true value of 4, and remains stable when $\Delta t\geq 4$. Hence, we should choose a reasonably large $\Delta t$ when its true value cannot be reliably inferred from the data or prior knowledge.

\section{Real-world application}
\label{sec:application}

This section presents the results of applying BOCD-AR to search interest data. The supplementary material provides the results of applying BOCD to the same data and the results of applying both algorithms to electric load data. Our search interest dataset contains hourly volumes of search queries related to stock prices in the United States. Early detection of collective anomalies and change points facilitates real-time analysis of trading behavioral changes in financial markets \citep{mavragani2018assessing, jun2018ten}. We collect the data from Google Trends through a Python package, pytrends \citep{googletrends2024, vazquez2020pytrends}. We exclude data on Fridays and weekends as their intraday patterns differ substantially from those on other days. The dataset includes 7,945 observations from January 1, 2023 through August 1, 2024. 

The experimental setup is as follows. We use Bayesian linear regression (see the supplementary material) with hyperparameters $\sigma_0^2=0.0001$, $v_0=0.01$, and $k_0=0.0001$ to model each data segment divided by changes. The independent variables include a day index and 24 binary variables indicating the hours of the day. Both independent and dependent variables are normalized using the min-max scaling. The hyperparameters of our method are set to $u_a=192$, $\Delta t=32$, $p_0=0.001$, $q_0=0.02$, $\lambda_a=0.5$, $\lambda_c=0.5$, and $\delta=6$. Instead of determining search time range $\Upsilon_c^t$ by $u_c$, we truncate it where the cumulative probability mass of $r^t$ is below $0.001$, subject to a minimum length of $1000$. A collective anomaly is confirmed only after observing at least $\Delta t=32$ subsequent data points, which prevents misclassifying changes recurring at specific times of day as multiple anomalies.

\begin{figure}[h]
	\centering
	\begin{subfigure}[b]{0.9\textwidth}
		\includegraphics[width=\textwidth]{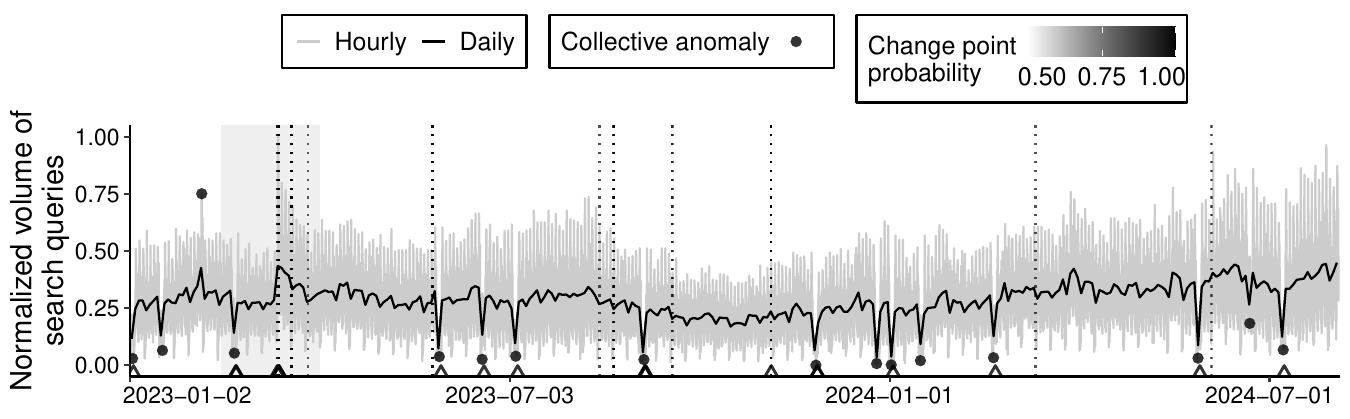}
	\caption{}
	\label{fig:data1_detect_all}
	\end{subfigure}
	\begin{subfigure}[b]{0.9\textwidth}
		\includegraphics[width=\textwidth]{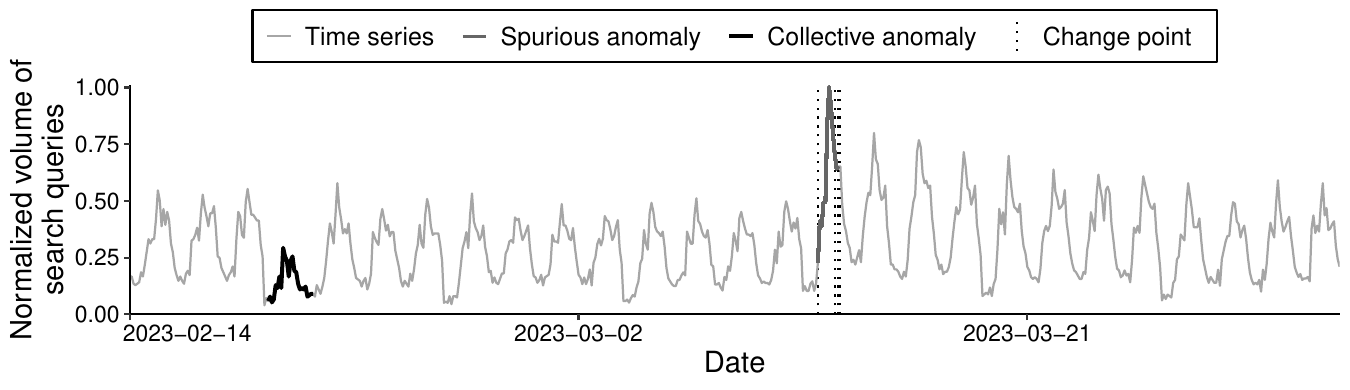}
	\caption{}
        \label{fig:data1_detect_short}
	\end{subfigure}
	\caption{Change detection results (a) for the entire study period and (b) from 2023-02-14 to 2023-03-30. The collective anomaly and change point in (b) are caused by Presidents' Day and the 2023 United States banking crisis \citep{bankingcrisis2023}, respectively.}
	\label{fig:data1_detect}
\end{figure}

Figure \ref{fig:data1_detect_all} illustrates the collective anomaly and change point detection result, and Figure \ref{fig:data1_detect_short} zooms into the shaded area. In Figure \ref{fig:data1_detect_all}, the grey line represents hourly volumes of search queries, and the black line indicates their daily averages. The black dots are the detected collective anomalies, mostly occurring on public holidays when financial markets are closed. The dotted vertical lines indicate the timings of the most recent change points that BOCD-AR detects at over 48 time points, with a darker color indicating a higher posterior probability. They align with events such as the 2023 United States banking crisis \citep{bankingcrisis2023} and the daylight saving time transitions. However, our current method cannot determine whether changes detected at different time points are related, a limitation we plan to address in future work. 

We compare our method with BOCPD, with the triangular points in Figure \ref{fig:data1_detect_all} indicating the timings of the most recent change points that BOCPD detects across over 48 time points. These changes generally align with the changes detected by our method. Nevertheless, BOCPD overlooks several evident collective anomalies and change points. The two methods exhibit similar detection delays. For collective anomalies, the average detection delays are 91.9 hours with our approach and 87.8 hours with BOCPD. Only one change point, which occurs on 2023-11-02, is detected by both methods and located far from anomalies. For this change point, our method reports a delay of 169 hours, while BOCPD reports 168 hours.

\section{Discussion}
\label{sec:discussion}

In this research, we propose a Bayesian online collective anomaly and change point detection method for fine-grained time series. We model change occurrences using two sets of binary variables, and set their prior probabilities to depend on the timing and type of the most recent change. We develop two recursive algorithms to update the posterior distributions of the most recent collective anomaly and change point. Experimental results demonstrate the effectiveness of our method. To the best of our knowledge, we are the first to develop a unified Bayesian framework that jointly detects both collective anomalies and change points in an online manner. Our method supports dynamic system monitoring in various fields, such as finance, biomedical sciences, and manufacturing. 

Despite its advantages, our method has limitations that should be addressed in future research. First, our algorithms ignore the similarity between data segments before and after a potential anomaly. This reduces the data utilization efficiency in distinguishing collective anomalies from spurious anomalies. A possible solution is to incorporate the penalized cost minimization approach of \cite{fisch2022linear}. Second, we assume that collective anomalies follow the same type of distribution as normal data. This assumption may not hold because anomalies result from abnormal system behavior. Future work could explore nonparametric techniques, such as kernel-based methods \citep{li2015m}, to overcome this limitation. Third, our method only infers the most recent change point. It cannot determine whether changes detected at different times refer to the same data shift, which limits its ability to uncover causal events. Future work could integrate the online multiple change point detection method of \cite{fearnhead2007line}.



\section{Supplementary material}
\label{sec:Supplementary}

The supplementary material includes (i) proofs of Theorems \ref{thm1}-\ref{thm3}, (ii) algorithmic overview and complexity analysis, (iii) hyperparameter choice analysis, (iv) description of Bayesian linear regression, (v) visualization of the change detection procedure, (vi) calculation of evaluation criteria, (vii) setup of an additional simulation, (viii) partial comparison of our algorithms, (ix) application of BOCD to search interest data, and (x) application of BOCD and BOCD-AR to electric load data. The code will be publicly available at Github.

\bibliographystyle{plainnat}
\bibliography{references}

\end{document}